\documentclass[prd,aps,tightenlines,a4paper,12pt]{revtex4}

\usepackage{graphicx}
\usepackage{bm}

\usepackage{url}

\newcommand{\ve}[1]{\mbox{\boldmath$#1$}}

\begin{document}

\title{Formal proof of some inequalities used in the analysis of the 
post-post-Newtonian light propagation theory}

\author{Sven \surname{Zschocke}, Sergei A. \surname{Klioner}}

\affiliation{Lohrmann Observatory, Dresden Technical University,
Mommsenstr. 13, 01062 Dresden, Germany}

\begin{abstract}
\begin{center}
{\it GAIA-CA-TN-LO-SZ-003-1}

\medskip

\today

\end{center}

A rigorous analytical solution of light propagation in Schwarzschild
metric in post-post Newtonian approximation has been presented in
\cite{report1}. In \cite{report2} it has been asserted that the sum of
all those terms which are of order $\displaystyle{{\cal O} \left(\frac{m^2}{d^2}\right)}$
and $\displaystyle{{\cal O}\left(\frac{m^2}{d_{\sigma}^2}\right)}$ is not greater than
$\displaystyle{\frac{15}{4}\,\pi\,\frac{m^2}{d^2}}$ and
$\displaystyle{\frac{15}{4}\,\pi\, \frac{m^2}{d_{\sigma}^2}}$,
respectively. All these terms can be neglected on microarcsecond level
of accuracy, leading to considerably simplified analytical
transformations of light propagation. In this report, we give formal
mathematical proofs for the inequalities used in the appendices of
\cite{report2}.

\end{abstract}


\maketitle

\newpage

\tableofcontents

\newpage

\section{Introduction}

In \cite{report1}, the rigorous analytical solution of light
propagation in Schwarzschild metric has been presented in post-post
Newtonian approximation. Especially, the analytical expressions for
Shapiro delay, the three transformations between $\ve{k}$ and
$\ve{\sigma}$, $\ve{\sigma}$ and $\ve{n}$, $\ve{k}$ and $\ve{n}$, and
the transformation between $\ve{n}$ and $\ve{\sigma}$ for stars and
quasars have been given explicitly. A detailed investigation in
\cite{report2} has shown that many of the terms, occurring in these five
transformations, are of order $\displaystyle{{\cal O} \left(\frac{m^2}{d^2}\right)}$ and
$\displaystyle{{\cal O} \left(\frac{m^2}{d_{\sigma}^2}\right)}$.  At the microarcsecond
level of accuracy all these terms can be neglected, leading to
considerably simplified analytical transformations. Furthermore, in
\cite{report2} it has been claimed that the sum of all these terms is
not greater than $\displaystyle{\frac{15}{4}\,\pi\,\frac{m^2}{d^2}}$
and $\displaystyle{\frac{15}{4}\,\pi\,\frac{m^2}{d_{\sigma}^2}}$,
respectively. This has been demonstrated without formal proof of
several inequalities involving functions $f_1, f_2, ..., f_{10}$ as
defined in the appendices of \cite{report2}.  The goal of this report
is to close this gap and to give formal proofs of these inequalities.
Throughout this report, for all estimations we use that $0 \le \Phi
\le \pi$, $0 \le \Psi \le \pi$ and $z \ge 0$; the angles $\Phi, \Psi$ 
and variable $z$ were defined in \cite{report2}.

\section{Estimate of function $f_1$}\label{f_1}

In this section we want to proof the inequality (A2) from Appendix A of \cite{report2}:
\begin{eqnarray}
f_1 &=& \frac{2\;z\;\left(1-\cos \Phi\right)}{1+z^2-2\,z\,\cos \Phi} \le \frac{4\,z}{\left(1+z\right)^2} \le 1\,.
\label{f1_5}
\end{eqnarray}

\noindent
The proof of the first inequality in (\ref{f1_5}) is straightforward and can be written as inequality
\begin{eqnarray}
- \left( 1 - z \right)^2\;\left(1 + \cos \Phi \right) &\le& 0\,,
\label{f1_10}
\end{eqnarray}

\noindent
which is obviously valid. The second inequality in (\ref{f1_5}) is equivalent to 
\begin{eqnarray}
4\,z - \left(1 + z \right)^2 &=& - \left( 1 - z \right)^2 \le 0\;.
\label{f1_15}
\end{eqnarray}

\noindent
From the inequalities (\ref{f1_10}) and (\ref{f1_15}) we conclude the validity of (\ref{f1_5}).

\section{Estimate of function $f_2$}\label{f_2}

In this section we want to proof the inequality (A4) from Appendix A of \cite{report2}:
\begin{eqnarray}
f_2 &=& 
\left|\,\sin \Phi\;\frac{z^2\,\cos \Phi - 2\,z + \cos \Phi}{1 + z^2 - 2\,z\,\cos \Phi} + 15\,\Phi\,\right| \le 
15\,\pi \,.
\label{f2_5}
\end{eqnarray}

To proof the validity of (\ref{f2_5}) we consider the extremal conditions
$f_{2,\Phi} = 0$ and $f_{2,z} = 0$, which yield
\begin{eqnarray}
0 &=&
7 - z^3\,\cos^3 \Phi - z\,\cos^3 \Phi + z^4\,\cos^2 \Phi + 32\,z^2\,\cos^2 \Phi
\nonumber\\
&& + \cos^2 \Phi - 31\, z^3\, \cos \Phi - 31\, z\, \cos \Phi + 7\,z^4 + 16\,z^2 \;,
\label{f2_10}
\\
0 &=& \sin^3 \Phi \; \left(z^2 - 1 \right).
\label{f2_15}
\end{eqnarray}

\noindent
Inserting the solutions of Eq.~(\ref{f2_15}), given by $\Phi = 0\;,\Phi = \pi\;,z = 1$,
into Eq.~(\ref{f2_10}) yields the equations
\begin{eqnarray}
0 &=& \left( 1 - z \right)^4 \,,
\label{f2_20}
\\
0 &=& \left( 1 + z \right)^4 \,,
\label{f2_25}
\\
0 &=& \left(\cos \Phi - 1 \right)^2 \left(\cos \Phi - 15 \right) \,.
\label{f2_30}
\end{eqnarray}

\noindent
The solution of Eq.~(\ref{f2_20}) is $z=1$, while Eq.~(\ref{f2_25}) has no solution, and the only solution 
of Eq.~(\ref{f2_30}) is $\Phi = 0$.  Thus, the extremal point $P_{\rm e}: \left(\Phi=0,z=1\right)$ 
which is only one special point of one of the boundaries of function $f_2$. The boundaries are given by
\begin{eqnarray}
f_2\,\bigg|_{z=0} &=& \sin \Phi\;\cos \Phi + 15\,\Phi \le 15\,\pi\,,
\label{f2_35}
\\
f_2\,\bigg|_{z = \infty} &=& \sin \Phi\,\cos \Phi + 15\,\Phi \le 15\,\pi\,,
\label{f2_40}
\\
f_2 \,\bigg|_{\Phi=0} &=& 0\,,
\label{f2_45}
\\
f_2 \,\bigg|_{\Phi=\pi} &=& 15\,\pi\,.
\label{f2_50}
\end{eqnarray}

\noindent
From Eqs.~(\ref{f2_35}) - (\ref{f2_50}) we conclude the validity of inequality (\ref{f2_5}).

\section{Estimate of function $f_3$}\label{f_3}

In this section we want to proof the inequality (B2) from Appendix B of \cite{report2}:
\begin{eqnarray}
f_3 = {1-z\over\sqrt{1+z^2-2z\cos\Phi}}+1
&\le&
\left[
\
\begin{array}{ll}
2\,,\quad&  z\le 1 \\[5pt] \displaystyle{\frac{2}{1 + z}}\,, & z>1
\end{array}
\right.\,
\le 2 \,.
\label{f3_5}
\end{eqnarray}

\noindent
Let us first consider the case $z \le 1$. In this case we square (\ref{f3_5}) and is equivalent to the inequality  
\begin{eqnarray}
1 + z^2 - 2 \,z \le 1 + z^2 - 2\,z\,\cos \Phi\,,
\label{f3_15}
\end{eqnarray}

\noindent
which is obviously valid. Let us now consider the  case $z > 1$, where we have to show 
\begin{eqnarray}
\frac{1-z}{\sqrt{1+z^2 - 2\,z\,\cos \Phi}} &\le& \frac{1-z}{1+z}\,,
\label{f3_20}
\end{eqnarray}

\noindent
or (note, that $1-z$ is negative)
\begin{eqnarray}
\sqrt{1+z^2 - 2\,z\,\cos \Phi} &\le& 1+z\,,
\label{f3_25}
\end{eqnarray}

\noindent
which is obviously valid. Thus, by means of (\ref{f3_15}) and (\ref{f3_25}), we have shown 
the validity of (\ref{f3_5}).

\section{Estimate of function $f_4$}\label{f_4}

In this section we want to proof the inequality (B4) from Appendix B of \cite{report2}:
\begin{eqnarray}
f_4 &=& z\,(1+z)\,{1-\cos\Phi\over 1+z^2-2z\,\cos\Phi}\,\left(1+{1-z\over\sqrt{1+z^2-2z\,\cos\Phi}}\right)
\nonumber\\
&&\le
\left[ \
\begin{array}{ll}
{16\over 27}\,(1+z),\ \  &  {1\over 2}\le z\le 1, \\[5pt]
\displaystyle{4\,\frac{z}{(1+z)^2}}, & z<{1\over 2}\ {\rm or}\ z> 1\,. \end{array}
\right.
\label{f4_5}
\end{eqnarray}

\subsection{$1/2 \le z \le 1$}

Let us first consider the case $1/2 \le z \le 1$, where (\ref{f4_5}) reduces to the inequality 
\begin{eqnarray}
\frac{16}{27} - \frac{z \left(1-w\right)}{1+z^2 - 2\,w\,z} &\ge&
\frac{z \left(1-w\right)\left(1-z\right)}{\left(1+z^2 - 2\,w\,z\right)^{3/2}} \,,
\label{f4_10}
\end{eqnarray}

\noindent
where $w=\cos \Phi$. Note, while the r.h.s. of (\ref{f4_10}) is obviously positive, 
the l.h.s. of (\ref{f4_10}) is also positive, because the inequality 
\begin{eqnarray}
\frac{16}{27} - \frac{z \left(1-w\right)}{1+z^2 - 2\,w\,z} &\ge& 0 
\label{f4_15}
\end{eqnarray}

\noindent
leads to $16+16\,z^2 -5\,w\,z -27\,z \ge 16 \left( 1 - z \right)^2 \ge 0$. 
Therefore, by squaring both sides of (\ref{f4_10}), we obtain the equivalent inequality 
\begin{eqnarray}
\left( 8 - 25\,w\,z + 9\,z + 8\,z^2 \right) \left(w \, z + 4 - 9\,z + 4\,z^2 \right)^2 &\ge& 0\,.
\label{f4_20}
\end{eqnarray}

\noindent
Since the quadratic term in (\ref{f4_20}) is by definition larger than zero, we have only to 
show that 
\begin{eqnarray}
h_1 &=& 8 - 25\,w\,z + 9\,z + 8\,z^2 \ge 0\,.
\label{f4_25}
\end{eqnarray}

\noindent
The extremal conditions $h_{1,w} = 0$ and $h_{1,z} = 0$ yield 
\begin{eqnarray}
-25\,z &=& 0\,,
\label{f4_30}
\\
-25\,w + 9 + 16\,z &=& 0\,.
\label{f4_35}
\end{eqnarray}

\noindent
The solution of (\ref{f4_30}) is $z=0$, however the region under consideration is $1/2 \le z \le 1$, 
that means there is no extremal point. The boundaries of function $h_1$ are 
\begin{eqnarray}
h_1 \, \bigg|_{z=1/2} &=& \frac{29}{2} - \frac{25}{2}\,w > 0 \,,
\label{f4_40}
\\
h_1\, \bigg|_{z=1} &=& 25 \left( 1 - w \right) \ge 0\,,
\label{f4_45}
\\
h_1\,\bigg|_{w=-1} &=& 8\,z^2 +34\,z + 8 > 0\,,
\label{f4_50}
\\
h_1\, \bigg|_{w=1} &=& 8\left( z - 1 \right)^2 \ge 0\,.
\label{f4_55}
\end{eqnarray}

\noindent
From Eqs.~(\ref{f4_40}) - (\ref{f4_55}) we conclude the validity of inequality (\ref{f4_25}) and 
(\ref{f4_10}). 

\subsection{$z> 1$}

The case $z> 1$ has actually been shown already, because by means of inequality (\ref{f1_5}) we obtain 
the estimate 
\begin{eqnarray}
f_4 &\le& \frac{2\,z}{\left(1+z\right)}\left(1 + \frac{1-z}{\sqrt{1+z^2 - 2\,z\,\cos \Phi}}\right),
\label{f4_60}
\end{eqnarray}

\noindent
and with the aid of inequality (\ref{f3_5}) we just obtain the estimate (\ref{f4_5}) for $z > 1$.

\subsection{$z < 1/2 $}

Let us now consider the case $z < 1/2$, where the inequality (\ref{f4_5}) reduces to 
\begin{eqnarray}
\frac{\left(1+z\right)^3 \left(1-w\right)}{1 + z^2 - 2\,w\,z} 
+ \frac{\left(1+z\right)^3 \left(1-w\right) \left(1-z\right)}{\left(1 + z^2 - 2\,w\,z\right)^{3/2}} &\le& 4\,.
\label{f4_65}
\end{eqnarray}

\noindent
We simplify (\ref{f4_65}) as follows:
\begin{eqnarray}
z^2 - z^2\,w - 4\,w\,z + 3 + w &\ge& \frac{\left(1+z\right)^3 \left(1-w\right)}{\sqrt{1 + z^2 - 2\,w\,z}} \,.
\label{f4_70}
\end{eqnarray}

\noindent
Squaring both sides of (\ref{f4_70}), which obviously are positive, yields the relation 
\begin{eqnarray}
&& \left( - z^5 - 8\,z^4 - 14\,z^3 + 8\,z^2 - z \right) w^2 
+ \left(4\,z^5 + 16\,z^4 + 8\,z^3 +16\,z^2 -12\,z \right) w 
\nonumber\\
&& - 3\,z^5 - 4\,z^4 - 10\,z^3 - 3\,z + 4 \ge 0\,.
\label{f4_75}
\end{eqnarray}

\noindent
In order to proof the validity of (\ref{f4_75}), we recall that $-1\le w \le 1$, that means 
the inequality (\ref{f4_75}) is valid, if the following inequality is satisfied:
\begin{eqnarray}
h_2 &=& \left|\,- z^5 - 8\,z^4 - 14\,z^3 + 8\,z^2 - z\,\right|
+ \left|\,4\,z^5 + 16\,z^4 + 8\,z^3 +16\,z^2 -12\,z\,\right| 
\nonumber\\
&& + \left|- 3\,z^5 - 4\,z^4 - 10\,z^3 - 3\,z\,\right| \le 4\,.
\label{f4_80}
\end{eqnarray}

\noindent
To proof inequality (\ref{f4_80}), we first note that 
\begin{eqnarray}
\left|\,4\,z^5 + 16\,z^4 + 8\,z^3 +16\,z^2 -12\,z\,\right| &\le& 
4\,z^5 + \left|\,16\,z^4 + 8\,z^3 +16\,z^2 -12\,z\,\right|\,.
\label{f4_85}
\end{eqnarray}

\noindent
Second, we note the obvious inequalities 
\begin{eqnarray}
- z^5 - 8\,z^4 - 14\,z^3 + 8\,z^2 - z = - z \left(z^2 + 4\,z - 1 \right)^2 &\le& 0\,,
\label{f4_90}
\\
16\,z^4 + 8\,z^3 +16\,z^2 -12\,z = 4\,z \left(2\,z -1 \right) \left(2\,z^2 + 2\,z + 3 \right) &\le& 0\,,
\label{f4_95}
\\
- 3\,z^5 - 4\,z^4 - 10\,z^3 - 3\,z\, &\le& 0\,.
\label{f4_100}
\end{eqnarray}

\noindent
Accordingly, by means of relation (\ref{f4_85}) and inequalities (\ref{f4_90}) - (\ref{f4_100}), 
we obtain 
\begin{eqnarray}
h_2 &\le& z \left(z^2 + 4\,z - 1 \right)^2 + 4\,z \left(1 - 2\,z \right) \left(2\,z^2 + 2\,z + 3 \right) 
\nonumber\\
&&
+ \left(3\,z^5 + 4\,z^4 + 10\,z^3 + 3\,z \right) + 4\,z^5 \le 4 \,.
\label{f4_102}
\end{eqnarray}

\noindent
Accordingly, we have to show the inequality 
\begin{eqnarray}
h_3 &=& z \; \left|\,2\,z^4 - z^3 + 4\,z^2 - 6\,z + 4 \, \right| \le 1 \,.
\label{f4_105}
\end{eqnarray}

\noindent
The extremal condition $h_{3,z} = 0$ yields 
\begin{eqnarray}
5\,z^4 - 2\,z^3 + 6\,z^2 - 6\,z + 2 &=& 0\,.
\label{f4_110}
\end{eqnarray}

\noindent
Eq.~(\ref{f4_110}) has no real solution due to 
$5\,z^4 - 2\,z^3 + 6\,z^2 - 6\,z + 2 \ge 2 \left( 1 - z \right)^3 > 0$ for 
$0 \le z \le 1/2$. Thus, there are no extremal points of $h_3$ in the region under consideration. The boundaries 
of function $h_3$ are given by 
\begin{eqnarray}
h_3\, \bigg|_{z=0} &=& 0\,,
\label{f4_115}
\\
h_3\, \bigg|_{z=1/2} &=& 1\,.
\label{f4_120}
\end{eqnarray}

\noindent
From (\ref{f4_115}) and (\ref{f4_120}) we conclude the validity of inequality (\ref{f4_105}), and 
by means of which we conclude the validity of inequality (\ref{f4_65}).

\section{Estimate of function $f_5$}\label{f_5}

In this section we want to proof the inequality (B6) from Appendix B of \cite{report2}:
\begin{eqnarray}
f_5 &=& 
\Bigg|\,-\frac{z (z^2 - 1) \sin^3\Phi}{\left(1 + z^2 - 2\,z\,\cos\Phi\right)^2} 
- 15\,\arccos \frac{1 - z\,\cos \Phi}{\sqrt{1 + z^2 - 2\,z\,\cos\Phi}}
\nonumber\\
&&
+15\,\frac{z\left(\cos\Phi- z\right)\Phi}{1 + z^2 - 2\,z\,\cos\Phi} + 15\,\pi\,\Bigg| \le 15\,\pi\,.
\label{f5_5}
\end{eqnarray}

\noindent
The extremal conditions $f_{5,\Phi}=0$ and $f_{5,z}=0$ yield
\begin{eqnarray}
0 &=& \sin \Phi\;z\;(z-1)\;(z+1)\;
\bigg(-3\,z^2\,\sin \Phi \, \cos \Phi + 15\,z^2\,\Phi + 2\,z\,\sin \,\Phi \cos^2 \Phi 
\nonumber\\
&& + 4\,z\,\sin \Phi - 30\,z\,\Phi\,\cos \Phi - 3\,\sin \Phi \,\cos \Phi + 15\,\Phi\bigg) \,,
\label{f5_10}
\\
0 &=& C_0 + C_1\;z + C_2\;z^2 + C_3\;z^3 + C_4\;z^4 \,,
\label{f5_15}
\end{eqnarray}

\noindent
with
\begin{eqnarray}
C_0 &=& C_4 = - \sin^3 \Phi + 15\;\sin \Phi - 15\;\Phi\;\cos \Phi\;,
\label{f5_20}
\\
C_1 &=& C_3 = 30\;\Phi + 30\;\Phi\;\cos^2 \Phi - 2\;\sin^3 \Phi \;\cos \Phi
- 60\;\sin \Phi \;\cos \Phi\;,
\label{f5_25}
\\
C_2 &=& - 90\;\;\Phi \;\cos \Phi +30\;\sin \Phi + 60\;\sin \Phi\;\cos^2 \Phi + 6\;\sin^3 \Phi\,.
\label{f5_30}
\end{eqnarray}

\noindent
The complete set of solutions of extremal condition $f_{5,\Phi}=0$ reads 
\begin{eqnarray}
z &=& 0\;,
\label{f5_35}
\\
z &=& 1\;,
\label{f5_40}
\\
\Phi &=& \pi\;,
\label{f5_45}
\\
\Phi &=& 0\;.
\label{f5_50}
\end{eqnarray}

\noindent
Note, that the expression in the parentheses of Eq.~(\ref{f5_10}) vanishes only at $\Phi=0$ (see Appendix 
\ref{appendixA}) which, however, represents no additional solution because it is already considered in 
Eq.~(\ref{f5_50}). Inserting the solutions (\ref{f5_35}) - (\ref{f5_45}) into (\ref{f5_15}) considerably 
simplifies the extremal condition $f_{5,z}=0$ and yields the relations  
\begin{eqnarray}
0 &=&  - \sin^3 \Phi + 15\,\sin \Phi - 15\, \Phi\,\cos \Phi \,,
\label{f5_55}
\\
0 &=& \left( 1 - \cos \Phi \right)^2\;
\left(15\;\Phi + \sin \Phi \left( 1 + \cos \Phi \right) +15\, \sin \Phi \right) \,,
\label{f5_60}
\\
0 &=& \left(z + 1\right)^4 \,,
\label{f5_65}
\end{eqnarray}

\noindent
while inserting the solution (\ref{f5_50}) into $f_{5,z} = 0$ yields an identity $0 = 0$, that means  
no additional solution. The only solution of (\ref{f5_55}) and (\ref{f5_60}) is given by 
(see Appendix \ref{appendixB})
\begin{eqnarray}
\Phi &=& 0\,,
\label{f5_70}
\end{eqnarray}

\noindent
while Eq.~(\ref{f5_65}) has obviously no solution.
In view of the solutions (\ref{f5_35}) - (\ref{f5_50}) and (\ref{f5_70})
we obtain the extremal points $P_{\rm e1}: \left(z = 0, \Phi = 0\right)$ and
$P_{\rm e2}: \left(z = 1, \Phi = 0\right)$. These extremal points are only two points on 
one of the boundaries. The boundaries are given by
\begin{eqnarray}
f_5\,\bigg|_{z=0} &=& 15\,\Phi \le 15\,\pi\,,
\label{f5_80}
\\
f_5\,\bigg|_{z=\infty} &=& \left|\,-15\,\pi + 15\,\Phi\,\right| \le 15\,\pi \,,
\label{f5_85}
\\
f_5\,\bigg|_{\Phi=0} &=& 15\,\pi \left(1 - \arccos \frac{1}{\sqrt{1+z^2}}\right) \le 15\,\pi\,,
\label{f5_90}
\\
f_5\,\bigg|_{\Phi=\pi} &=& \frac{15\,\pi}{1+z} \le 15\,\pi\,.
\label{f5_100}
\end{eqnarray}

\noindent
From Eqs.~(\ref{f5_80}) - (\ref{f5_100}) we conclude the validity of inequality (\ref{f5_5}).

\section{Estimate of function $f_6$}\label{f_6}

In this section we want to proof the inequality (C2) from Appendix C of \cite{report2}:
\begin{eqnarray}
f_6 &=& \left({1-z\,\cos\Phi\over \sqrt{1+z^2-2z\,\cos\Phi}}+1\right)\,
{z\,(1-\cos\Phi)\over 1+z^2-2z\,\cos\Phi}\le{4z\over(1+z)^2}\le 1\,. 
\label{f6_5}
\end{eqnarray}

\noindent
The second inequality has been shown in Eq.~(\ref{f1_15}), thus we focus on the first inequality only. 
By inserting (\ref{f1_5}) into first inequality of (\ref{f6_5}), we recognize that we have only 
to show the considerably simpler inequality 
\begin{eqnarray}
\left| \, \frac{1-z\,\cos\Phi}{\sqrt{1+z^2-2z\,\cos\Phi}} \,\right| \,\le 1\,,
\label{f6_10}
\end{eqnarray}

\noindent
or 
\begin{eqnarray}
\left( 1 - z\,\cos \Phi \right)^2 &\le& 1 + z^2 - 2\,z\,\cos \Phi\,.
\label{f6_15}
\end{eqnarray}

\noindent
The inequality (\ref{f6_15}) is, however, obviously valid. Thus we have shown the validity of 
first inequality of (\ref{f6_5}).

\section{Estimate of function $f_7$}\label{f_7}

In this section we want to proof the inequality (C4) from Appendix C of \cite{report2}:
\begin{eqnarray}
f_7 &=& \left|\,16\,\sin\Psi+\sin\Psi\,\cos\Psi-2\sin^3\Psi\,\cos\Psi-15\,\pi+15\,\Psi\,\right| \le 15\,\pi \,.
\label{f7_5}
\end{eqnarray}

\noindent
The extremal condition $f_{7,\Psi} = 0$ yields 
\begin{eqnarray}
\left( 1 + \cos \Psi \right)^2 \left( \cos^2 \Psi - 2\,\cos \Psi + 2 \right)   &=& 0\,.
\label{f7_15}
\end{eqnarray}
 
\noindent
The only solution of Eq.~(\ref{f7_15}) is $\Psi = \pi$ that means the extremal point is 
$P_{\rm e}: \left(\Psi = \pi \right)$, which is basically the boundary given below in Eq.~(\ref{f7_40}). 
The boundaries are given by 
\begin{eqnarray}
f_7\,\bigg|_{\Psi=0} &=&  15\,\pi\,,
\label{f7_35}
\\
f_7\,\bigg|_{\Psi=\pi} &=& 0\,.
\label{f7_40}
\end{eqnarray}

\noindent
From Eqs.~(\ref{f7_35}) and (\ref{f7_40}) we conclude the validity of (\ref{f7_5}).

\section{Estimate of function $f_8$}\label{f_8}

In this section we want to proof the inequality (D2) from Appendix D of \cite{report2}:
\begin{eqnarray}
f_8 &=& {z\,(1-\cos\Phi)\over \sqrt{1+z^2-2z\,\cos\Phi}}\le {2z\over 1+z}\,,
\label{f8_5}
\end{eqnarray}

\noindent
which is equivalent to the inequality  
\begin{eqnarray}
h_4 &=& w + 2\,w\,z + z^2\,w - 3 + 2\,z - 3\,z^2 \le 0\,.
\label{f8_15}
\end{eqnarray}

\noindent
The extremal conditions $h_{4,z}=0$ and $h_{4,w}=0$ yield
\begin{eqnarray}
w + w\,z +1 - 3\,z &=& 0\,,
\label{f8_20}
\\
\left( 1 + z \right)^2 &=& 0\,.
\label{f8_25}
\end{eqnarray}

\noindent
Eq.~(\ref{f8_25}) has obviously no real solution for variable $z \ge 0$, and, therefore, the function
$h_4$ has no extremal points. The boundaries of $h_4$ are given by
\begin{eqnarray}
h_4\,\bigg|_{z=0} &=&  w - 3 \le 0\,,
\label{f8_30}
\\
h_4\,\bigg|_{z=\infty} &=&  \left( w - 3 \right) \lim_{z \rightarrow \infty} z^2 \le 0\,,
\label{f8_35}
\\
h_4\,\bigg|_{w=-1} &=&  - 4\, \left(1 + z\right)^2 \le 0\,,
\label{f8_40}
\\
h_4\,\bigg|_{w=1} &=&  -2 \left( 1 - z \right)^2 \le 0\,.
\label{f8_45}
\end{eqnarray}

\noindent
From Eqs.~(\ref{f8_30}) - (\ref{f8_45}) we conclude the validity of (\ref{f8_15}) and (\ref{f8_5}).

\section{Estimate of function $f_9$}\label{f_9}

In this section we want to proof the inequality (D4) from Appendix D of \cite{report2}:
\begin{eqnarray}
f_9 &=& {z^2\,(1+z)\,{(1-\cos\Phi)}^2\over {(1+z^2-2z\,\cos\Phi)}^2}\le{4z^2\over (1+z)^3}\,,
\label{f9_5}
\end{eqnarray}

\noindent
which is equivalent to the inequality 
\begin{eqnarray}
h_5 &=& z^2 \,w - 3\,z^2 + 6\,w\,z - 2\,z + w - 3 \le 0\,,
\label{f9_15}
\end{eqnarray}

\noindent
where $w = \cos \Phi$. The extremal conditions $h_{5,z}=0$ and $h_{5,w}=0$ yield
\begin{eqnarray}
w\,z - 3\,z + 3\,w - 1 &=& 0\,,
\label{f9_20}
\\
z^2 + 6\,z + 1 &=& 0\,.
\label{f9_25}
\end{eqnarray}

\noindent
Eq.~(\ref{f9_25}) has obviously no real solution for variable $z \ge 0$, and, therefore, the function 
$h_5$ has no extremal points. The boundaries of $h_5$ are given by 
\begin{eqnarray}
h_5\,\bigg|_{z=0} &=& w - 3 \le 0\,,
\label{f9_30}
\\
h_5\,\bigg|_{z=\infty} &=& \left(w - 3 \right) \lim_{z \rightarrow \infty} z^2 \le 0\,,
\label{f9_35}
\\
h_5\,\bigg|_{w=-1} &=& - 4 \left(1+z\right)^2 \le 0\,,
\label{f9_40}
\\
h_5\,\bigg|_{w=1} &=& - 2 \left(1-z\right)^2 \le 0\,.
\label{f9_45}
\end{eqnarray}

\noindent
From Eqs.~(\ref{f9_30}) - (\ref{f9_45}) we conclude the validity of (\ref{f9_15}) and (\ref{f9_5}).

\section{Estimate of function $f_{10}$}\label{f_10}

In this section we want to proof the inequality (D6) from Appendix D of \cite{report2}:
\begin{eqnarray}
f_{10} &=& \Bigg|
{z\,(16z-z\,\cos\Phi-15)\,\sin\Phi\over 1+z^2-2z\,\cos\Phi}
+{z(1-3z^2+2z^3\cos\Phi)\,\sin^3\Phi\over \left(1+z^2-2z\,\cos\Phi\right)^2}
+{15z\,(\cos\Phi-z)\,\Phi\over 1+z^2-2z\,\cos\Phi}
\Bigg|
\nonumber\\
&\le& 15\,\pi\,.
\label{f10_5}
\end{eqnarray}

\noindent
With $\left| a + b \right| \le \left| a \right| + \left| b \right|$, and the inequality 
(see Appendix \ref{appendixC}) 
\begin{eqnarray}
\left|
\frac{z(1-3z^2+2z^3\cos\Phi)\,\sin^3\Phi}{\left(1+z^2-2z\,\cos\Phi\right)^2}\right|
&\le& 8\,\sin \Phi \,,
\label{f10_10}
\end{eqnarray}

\noindent
we get
\begin{eqnarray}
f_{10} &\le& \left|{z\,(16z-z\,\cos\Phi-15)\,\sin\Phi + 15z\,(\cos\Phi-z)\,\Phi\over 1+z^2-2z\,\cos\Phi}\right| 
+ 8\,\sin \Phi\,.
\label{f10_15}
\end{eqnarray}

\noindent
Due to the inequality (see Appendix \ref{appendixD})
\begin{eqnarray}
z \left( 16\,z - z \cos \Phi - 15 \right) \sin \Phi 
+ 15\,z\, \left( \cos \Phi - z \right) \Phi &\le& 0\,,
\label{f10_20}
\end{eqnarray}

\noindent
and since $\sin \Phi \ge 0$, we can write
\begin{eqnarray}
f_{10} &\le&
\left| \sin \Phi\;\frac{16 z^2 - 15 z - z^2 \cos \Phi}{1+z^2-2 z \cos \Phi}
- 8\, \sin \Phi
+ 15 \; \frac{z\;\Phi\;\left( \cos \Phi - z \right)}{1+z^2-2 z \cos \Phi} \right| \,.
\label{f10_25}
\end{eqnarray}

\noindent
Since the expression in the parentheses of Eq.~(\ref{f10_25}) is negative, we can replace $\cos \Phi$ by $1$ 
in the nominator of the first term (note, that $1 + z^2 - 2\,z\,\cos \Phi \ge 0$) and obtain
\begin{eqnarray}
f_{10} &\le& 
\left| 15\;\sin \Phi\;\frac{z \, \left( z - 1 \right)}{1+z^2-2 z \cos \Phi}
- 8 \; \sin \Phi
+ 15 \; \frac{z\;\Phi\;\left( \cos \Phi - z \right)}{1+z^2-2 z \cos \Phi} \right| \,.
\label{f10_30}
\end{eqnarray}

\noindent
This expression can further be simplified by means of the inequality (see Appendix \ref{appendixE})
\begin{eqnarray}
\left|\;\frac{z\;\Phi\;\left( \cos \Phi - z \right)}{1+z^2-2 z \cos \Phi} \;\right|
&\le& \frac{z\;\pi}{1+z} \,.
\label{f10_35}
\end{eqnarray}

\noindent
Thus we obtain
\begin{eqnarray}
f_{10} &\le& 
\left| 15\; \sin \Phi\;\frac{z\,\left( z - 1 \right)}{1+z^2-2 z \cos \Phi}
- 15 \,\sin \Phi - 15\; \frac{z\;\pi}{1+z}\; \right|\,,
\label{f10_40}
\end{eqnarray}

\noindent
where we also made the replacement $- 8\, \sin \Phi$ by $-15\,\sin \Phi$
for getting an expression more convenient for subsequent considerations.
This expression can simplified with the aid of the inequality (see Appendix \ref{appendixG})
\begin{eqnarray}
\left|\,\sin \Phi\;\frac{z\,\left( z - 1 \right)}{1+z^2-2 z \cos \Phi} - \sin \Phi \,\right|
&\le& \frac{2}{1 + z}\;,
\label{f10_45}
\end{eqnarray}

\noindent
by means of which we obtain
\begin{eqnarray}
f_{10} &\le& 15\,\left|\,\frac{2}{1 + z} + \frac{z\;\pi}{1 + z}\,\right|
\le  15\,\left|\,\frac{\pi}{1 + z} + \frac{z\;\pi}{1 + z}\,\right|
= 15 \, \pi \,.
\label{f10_50}
\end{eqnarray}

\noindent
Thus, we have shown the validity of inequality (\ref{f10_5}).

\newpage

\appendix

\section{Proof the non-existence of other solutions of Eq.~(\ref{f5_10})}\label{appendixA}

If we set the parentheses of Eq.~(\ref{f5_10}) to zero, we obtain the following equation:
\begin{eqnarray}
0 &=& -3\,z^2\,\sin \Phi \, \cos \Phi + 15\,z^2\,\Phi + 2\,z\,\sin \,\Phi \cos^2 \Phi
\nonumber\\
&& + 4\,z\,\sin \Phi - 30\,z\,\Phi\,\cos \Phi - 3\,\sin \Phi \,\cos \Phi + 15\,\Phi\,.
\label{appendix_f5_5}
\end{eqnarray}

\noindent
We want to show that Eq.~(\ref{appendix_f5_5}) has the only solution $P_{\rm e}: \left(z = 1, \Phi = 0\right)$.
The both solutions of Eq.~(\ref{appendix_f5_5}) for variable $z$ are given by
\begin{eqnarray}
z_{1,2} &=& \frac{1}{3}\;\frac{\sin^3 \Phi - 3\,\sin \Phi + 15\,\Phi\,\cos \Phi \pm \sqrt{T_1}}
{5\,\Phi - \sin \Phi\;\cos \Phi}\,,
\label{appendix_f5_10}
\end{eqnarray}

\noindent
where the discriminant is defined by
\begin{eqnarray}
T_1 &=& - \sin^2 \Phi \left( - \cos^4 \Phi + 5\,\cos^2 \Phi - 30\,\Phi\;\sin \Phi \; \cos \Phi
+ 225\;\Phi^2  - 4 \right) \le 0 \;.
\label{appendix_f5_15}
\end{eqnarray}

\noindent
The inequality (\ref{appendix_f5_15}) can also be expressed by
\begin{eqnarray}
h_6 &=& - w^4 + 5\;w^2 - 30\;\sqrt{1-w^2}\;w\;\arccos w + 225\;\arccos^2 w - 4 \ge 0 \;,
\label{appendix_f5_20}
\end{eqnarray}

\noindent
where $w = \cos \Phi$. The extremal condition $h_{6,w} = 0$ leads to
\begin{eqnarray}
0 &=& 120\;\arccos w - 10\;w\;\sqrt{1-w^2} - 15\;w^2\;\arccos w + w^3\;\sqrt{1-w^2}\,,
\label{appendix_f5_25}
\end{eqnarray}

\noindent
with the only solution $w=1$, that means $\Phi=0$ (it is straightforward to show, that 
the first derivative of expression (\ref{appendix_f5_25}) is always negative; thus the 
expression (\ref{appendix_f5_25}) represents a monotonically decreasing function
and since $w=1$ is obviously a solution of equation (\ref{appendix_f5_25}) it is,
therefore, the only one).  Inserting $\Phi=0$ into Eq.~(\ref{appendix_f5_10})
yields $z_1 = z_2 = 1$. Thus, the extremal point is given by 
\begin{eqnarray}
P_{\rm e}: (z=1, w = 1)\,.
\label{appendix_f5_30}
\end{eqnarray}

\noindent
The boundaries of function $h_6$ are given by
\begin{eqnarray}
h_6\,\bigg|_{w=-1} &=& 225\;\pi^2 > 0\,,
\label{appendix_f5_40}
\\
h_6,\bigg|_{w=1} &=& 0\,.
\label{appendix_f5_45}
\end{eqnarray}

\noindent
From Eqs.~(\ref{appendix_f5_40}) and (\ref{appendix_f5_45}) we conclude the validity of
inequality (\ref{appendix_f5_20}) and (\ref{appendix_f5_15}).
That means, the only real solution of Eq.~(\ref{appendix_f5_5}) is given by
Eq.~(\ref{appendix_f5_30}), that means $P_{\rm e}: \left(z=1, \Phi=0\right)$.

\section{Proof that Eq.~(\ref{f5_70}) is the only solution of Eq.~(\ref{f5_55}) and Eq.~(\ref{f5_60})}\label{appendixB}

Eq.~(\ref{f5_55}) is given by
\begin{eqnarray}
0 &=&  - \sin^3 \Phi + 15\,\sin \Phi - 15\, \Phi\,\cos \Phi \;,
\label{appendixB_5}
\end{eqnarray}

\noindent
and Eq.~(\ref{f5_60}) is given by
\begin{eqnarray}
0 &=& \left( 1 - \cos \Phi \right)^2\;
\left(15\;\Phi + \sin \Phi \left( 1 + \cos \Phi \right) +15\, \sin \Phi \right) \,.
\label{appendixB_10}
\end{eqnarray}

\noindent
The only solution of (\ref{appendixB_10}) is (it is straightforward to show that the first derivative of 
$15\;\Phi + \sin \Phi \left( 1 + \cos \Phi \right) +15\, \sin \Phi$ is always positive, that means this 
expression represents a monotonically increasing function; thus (\ref{appendixB_15}) is, 
therefore, the only solution of (\ref{appendixB_10}))
\begin{eqnarray}
\Phi &=& 0\;.
\label{appendixB_15}
\end{eqnarray}

\noindent 
Inserting the solution (\ref{appendixB_15}) into Eq.~(\ref{appendixB_5}) yields an identity $0=0$. Thus, 
the only solution of Eqs.~(\ref{appendixB_5}) and (\ref{appendixB_10}) is given by (\ref{appendixB_15}).

\section{Proof of inequality (\ref{f10_10})}\label{appendixC}

The inequality in Eq.~(\ref{f10_10}) is given by 
\begin{eqnarray}
\frac{\sin^3 \Phi}{\left(1+z^2-2\,z\,\cos \Phi\right)^2}
\left| z - 3\,z^3 + 2\,z^4 \cos \Phi \right| &\le& 8\, \sin \Phi\;,
\label{appendixC_5}
\end{eqnarray}

\noindent
which can be written as 
\begin{eqnarray}
\frac{1 - w^2}{\left(1+z^2-2\,w\,z\,\right)^2}
\left| \, z - 3\,z^3 + 2\,w\,z^4 \,\right| &\le& 8\;,
\label{appendixC_10}
\end{eqnarray}

\noindent
where $w = \cos \Phi$. 
In order to show the validity of Eq.~(\ref{appendixC_10}) it is convenient to simplify this expression 
with the aid of the following inequality:
\begin{eqnarray}
\frac{1 - w^2}{\left(1+z^2-2\,w\,z\,\right)^2}
\left| \, z - 3\,z^3 + 2\,w\,z^4 \,\right| &\le& 
2\; \frac{1 - w}{\left(1+z^2-2\,w\,z\,\right)^2}
\left| \, z - 3\,w\,z^3 + 2\,z^4 \,\right|\,.
\label{appendixC_15}
\end{eqnarray}

\noindent
The proof of inequality (\ref{appendixC_15}) can be shown as follows. 
Due to $1 - w^2 \le 2 \left( 1 - w \right)$ for $-1 \le w \le 1$, we have only to show 
$\left|\,z - 3\,z^3 + 2\,w\,z^4\,\right| \le \left|\,z - 3\,w\,z^3 + 2\,z^4\,\right|$.
Squaring both of these sides and subtracting from each other leads to the inequality 
\begin{eqnarray}
z^4 \left(w - 1 \right)\;\left(3 + 2\,z \right) 
\left( 2\,w\,z^3 - 3\,w\,z^2 + 2 + 2\,z^3 - 3\,z^2 \right) &\le& 0 \,.
\label{appendixC_20}
\end{eqnarray}

\noindent
That means, in order to proof (\ref{appendixC_15}), one has actually to show the inequality 
\begin{eqnarray}
g &=& 2\,w\,z^3 - 3\,w\,z^2 + 2 + 2\,z^3 - 3\,z^2 \ge 0\,.
\label{appendixC_25}
\end{eqnarray}

\noindent
The extremal point is given by $P_{\rm e} \left( z = 3/2, w=-1 \right)$ where 
$g\,|_{P_{\rm e}} = 2$ , and the boundaries are $g\,|_{z=0} = 2 \ge 0$, 
$g\,|_{z=\infty} = 2 \left( 1 + w\right) z^3 \ge 0$, 
$g\,|_{w=-1} = 2$, $g\,|_{w=1} =  4\,z^3 - 6\,z^2 + 2 \ge 0$.
Thus, we have shown the validity of (\ref{appendixC_25}) and (\ref{appendixC_15}), respectively.
Let us turn back to the original inequality (\ref{appendixC_5}),
\begin{eqnarray}
h_7 &=& \frac{1 - w}{\left(1+z^2-2\,w\,z\,\right)^2}
\left| \, z - 3\,w\,z^3 + 2\,z^4 \,\right| \le 4 \,,
\label{appendixC_30}
\end{eqnarray}

\noindent
which follows from the combination of (\ref{appendixC_10}) and (\ref{appendixC_15}). In order to show the 
validity of (\ref{appendixC_30}), we consider the extremal conditions $h_{7,z}=0$ and $h_{7,w}=0$ which yield 
\begin{eqnarray}
\left( 1 - w \right)\,\left(1 + 2\,w\,z - 3\,z^2 -9\,w\,z^2 + 8\,z^3 + 6\,w^2\,z^3 - 5\,w\,z^4 \right) &=& 0 \,,
\label{appendixC_35}
\\
z\;\left(z - 1 \right)^3 \, \left( - 2\,z^2 - z + 2\,w\,z + 1 \right) &=& 0\,.
\label{appendixC_40}
\end{eqnarray}

\noindent
The solutions of (\ref{appendixC_40}) are given by 
\begin{eqnarray}
z_1 &=& 0\,,
\label{appendixC_45}
\\
z_2 &=& 1 \,,
\label{appendixC_50}
\\
z_{3,4} &=& \frac{2\,w - 1}{4} \pm \frac{\sqrt{9 - 4\,w + 4\,w^2}}{4}\,,
\label{appendixC_55}
\\
w &=& \frac{2\,z^2 + z - 1}{2\,z}\,.
\label{appendixC_60}
\end{eqnarray}

\noindent
Inserting (\ref{appendixC_45}) - (\ref{appendixC_60}) into extremal condition (\ref{appendixC_35}) yields 
the relations 
\begin{eqnarray}
1 - w &=& 0\,,
\label{appendixC_65}
\\
\left(1 - w\right)^3 &=& 0\,,
\label{appendixC_70}
\\
\left(7\,w + 20\right)\;\left(4\,w + 5 \right)\;\left(w - 1 \right)^4 &=& 0 \,,
\label{appendixC_75}
\\
\left( z - 1 \right)^3 \;\left(z + 2\right)\;\left(2\,z + 1 \right)\;\left(2\,z + 7\right) &=& 0\,.
\label{appendixC_80}
\end{eqnarray}

\noindent
The relevant solutions of 
(\ref{appendixC_65}) - (\ref{appendixC_80}), respectively, are given by 
\begin{eqnarray}
w &=& 1\,,
\label{appendixC_85}
\\
z &=& 1 \,.
\label{appendixC_90}
\end{eqnarray}

\noindent
Thus, the extremal point is given by $P_{\rm e}: \left(z=1,w=1\right)$, which is just one 
point of one of the boundaries. The boundaries of $h_7$ are given by 
\begin{eqnarray}
h_7\,\bigg|_{z=0} &=& 0 \,,
\label{appendixC_100}
\\
h_7\,\bigg|_{z=\infty} &=& 2 \; \left( 1 - w \right) \le 4 \,,
\label{appendixC_105}
\\
h_7\,\bigg|_{w=-1} &=& 2\;z\;\frac{1 + 3\,z^2 + 2\,z^3}{\left(1 + z\right)^4} \le 4\,,
\label{appendixC_110}
\\
h_7\,\bigg|_{w=1} &=& 0 \,.
\label{appendixC_115}
\end{eqnarray}

\noindent
From Eqs.~(\ref{appendixC_100}) - (\ref{appendixC_115}) we conclude the validity of 
inequality (\ref{appendixC_30}), that means the validity of (\ref{appendixC_5}) and 
(\ref{appendixC_10}), respectively.

\section{Proof of inequality (\ref{f10_20})}\label{appendixD}

Since $z \ge 0$, Eq.~(\ref{f10_20}) can also be written by
\begin{eqnarray}
h_8 &=& \sin \Phi \left( 16\,z - 15 - z \cos \Phi \right)
+ 15\,\left( \cos \Phi - z \right) \Phi \le 0\,.
\label{appendixD_5} 
\end{eqnarray}

\noindent
The extremal conditions $h_{8,z}=0$ and $h_{8,\Phi}=0$ yield
\begin{eqnarray}
0 &=& 16\,\sin \Phi - \sin \Phi\;\cos \Phi - 15\;\Phi\,,
\label{appendixD_10}
\\
0 &=& 16\;z\;\cos \Phi - z\;\cos^2 \Phi + z\;\sin^2 \Phi - 15\;z - 15\;\Phi\;\sin \Phi\,.
\label{appendixD_15}
\end{eqnarray}

\noindent
The only solution of Eq.~(\ref{appendixD_10}) is given by (note, that it is straightforward to
show that the first derivative of (\ref{appendixD_5}) is negative, i.e. (\ref{appendixD_5})
represents a monotonically decreasing function; thus the given solution
(\ref{appendixD_20}) is indeed the only possible solution)
\begin{eqnarray}
\Phi &=& 0\,.
\label{appendixD_20}
\end{eqnarray}

\noindent
Inserting the solution (\ref{appendixD_20}) into (\ref{appendixD_15}) yields an identity $0=0$.
That means the function $h_8$ does not have an extremal point, while $h_8$ takes extremal values at the
boundary $\Phi=0$. Especially, the boundaries are given by
\begin{eqnarray}
h_8\,\bigg|_{z=0} &=& -15\,\sin \Phi +15\,\Phi\,\cos \Phi \le 0\,,
\label{appendixD_25}
\\
h_8\,\bigg|_{z=\infty} &=& 
\lim_{z \rightarrow \infty} \left( -15\,\Phi - \sin \Phi\, \cos \Phi +16\,\cos \Phi \right) z \le 0 \,,
\label{appendixD_30}
\\
h_8\,\bigg|_{\Phi=0} &=& 0\,,
\label{appendixD_35}
\\
h_8\,\bigg|_{\Phi=\pi} &=& -15 \, \pi \left( 1 + z \right) \le 0\,.
\label{appendixD_40}
\end{eqnarray}

\noindent
From Eqs.~(\ref{appendixD_25}) - (\ref{appendixD_40}) we conclude the validity of
inequality (\ref{appendixD_5}).

\section{Proof of inequality (\ref{f10_35})}\label{appendixE}

Eq.~(\ref{f10_35}) can be written by
\begin{eqnarray}
\left|\;\frac{\Phi\;\left( \cos \Phi - z \right)}{1 + z^2 - 2 \, z \, \cos \Phi} \;\right|
&\le& \frac{\pi}{1 + z} \,,
\label{appendixE_5}
\end{eqnarray}

\noindent
that means we have to show the validity of 
\begin{eqnarray}
\frac{\Phi^2\;\left( \cos \Phi - z \right)^2}{\left(1 + z^2 - 2 \, z \,\cos \Phi\right)^2} 
- \frac{\pi^2}{\left( 1 + z \right)^2} \le 0 \,.
\label{appendixE_10}
\end{eqnarray}

\noindent
Obviously, the following inequality is valid:
\begin{eqnarray}
\frac{\left( \cos \Phi - z \right)^2}{\left( 1 + z^2 - 2 \, z \, \cos \Phi\right)} &\le& 1 \,.
\label{appendixE_15}
\end{eqnarray}

\noindent
Thus, that means by inserting (\ref{appendixE_15}) into (\ref{appendixE_10}), we have to show the inequality 
\begin{eqnarray}
h_9 &=& \frac{\Phi^2}{1 + z^2 - 2 \, z \, \cos \Phi}
- \frac{\pi^2}{\left( 1 + z \right)^2} \le 0 \,.
\label{appendixE_20}
\end{eqnarray}

\noindent
The extremal conditions $h_{9,z}=0$ and $h_{9,\Phi}=0$ lead to 
\begin{eqnarray}
0 &=& -z\,\Phi - 2\,z^2\,\Phi - z^3\,\Phi + \Phi\,\cos \Phi + 2\,z\,\Phi\,\cos \Phi + z^2\,\,\Phi\,\cos \Phi 
\nonumber\\
&& + \pi \left( 1 + z^2 - 2\,z\,\cos \Phi \right)^{3/2},
\label{appendixE_25}
\\
0 &=& 1 + z^2 - 2\,z\,\cos \Phi - z\,\Phi\,\sin \Phi\,.
\label{appendixE_30}
\end{eqnarray}

\noindent
The relation (\ref{appendixE_30}) represents an quadratic equation in variable $z$ and has the both solutions 
\begin{eqnarray}
z_{1,2} &=& \cos \Phi + \frac{1}{2}\;\sin \Phi \pm \frac{1}{2}\;\sqrt{T_2}\;,
\label{appendixE_35}
\end{eqnarray}

\noindent
where the discriminant $T_2$ is given by 
\begin{eqnarray}
T_2 &=& \sin \Phi \left(-4\,\sin \Phi + 4\,\Phi \,\cos \Phi + \Phi^2\;\sin \Phi \right) \le 0 \,,
\label{appendixE_40}
\end{eqnarray}

\noindent
where the inequality (\ref{appendixE_40}) is shown in Appendix \ref{appendixF}. 
The discriminant $T_2 = 0$ at $\Phi=0$. Thus, in view of Eq.~(\ref{appendixE_35}) and (\ref{appendixE_40}), 
the only real solution of Eq.~(\ref{appendixE_30}) is given by $P \left(z=1, \Phi = 0\right)$. Inserting 
this solution into Eq.~(\ref{appendixE_25}) yields 
\begin{eqnarray}
0 &=& \pi\;\left(z - 1 \right)^3 .
\label{appendixE_45}
\end{eqnarray}

\noindent
Thus, the extremal point is given by 
\begin{eqnarray}
P_{\rm e}: \left(z=1, \Phi = 0\right),
\label{appendixE_50}
\end{eqnarray}

\noindent
which is just one point on one of the boundaries. The boundaries of function $h_9$ are given by 
\begin{eqnarray}
h_9\,\bigg|_{z=0} &=& \Phi^2 - \pi^2  \le 0\,,
\label{appendixE_60}
\\
h_9\,\bigg|_{z=\infty} &=& 0\,,
\label{appendixE_65}
\\
h_9\,\bigg|_{\Phi=0} &=& - \frac{\pi^2}{\left(1 + z\right)^2} \le 0\,,
\label{appendixE_70}
\\
h_9\,\bigg|_{\Phi=\pi} &=& 0\,.
\label{appendixE_75}
\end{eqnarray}

\noindent
From Eqs.~(\ref{appendixE_60}) - (\ref{appendixE_75}) we conclude the validity of 
inequality (\ref{appendixE_20}) and (\ref{appendixE_5}), respectively.

\section{Proof of inequality (\ref{appendixE_40})}\label{appendixF}

Inequality (\ref{appendixE_40}) is given by
\begin{eqnarray}
T_2 &=& \sin \Phi\;\left(-4\,\sin \Phi + 4\,\Phi \,\cos \Phi + \Phi^2\;\sin \Phi \right) \le 0 \,,
\label{appendixF_5}
\end{eqnarray}

\noindent
that means, due to $\sin \Phi \ge 0$, we have to show the inequality
\begin{eqnarray}
h_{10} &=& -4\,\sin \Phi + 4\,\Phi \,\cos \Phi + \Phi^2\;\sin \Phi \le 0 \,.
\label{appendixF_10}
\end{eqnarray}

\noindent
The extremal condition $h_{10,\Phi} = 0$ leads to
\begin{eqnarray}
0 &=& \Phi \left( \Phi \,\cos \Phi - 2\,\sin \Phi \right) ,
\label{appendixF_15}
\end{eqnarray}

\noindent
with the only solution (the first derivative of (\ref{appendixF_15}) is always negative, thus the expression 
on the r.h.s. of Eq.~(\ref{appendixF_15}) represents a monotonically decreasing function and, therefore, the 
given solution (\ref{appendixF_20}) is in fact the only possible solution)
\begin{eqnarray}
P_{\rm e}: \left( \Phi = 0\right).
\label{appendixF_20}
\end{eqnarray}

\noindent
The extremal point (\ref{appendixF_20}) is just one point on one of the both boundaries.  
The boundaries are given by
\begin{eqnarray}
h_{10}\,\bigg|_{\Phi=0} &=& 0 \,,
\label{appendixF_25}
\\
h_{10}\,\bigg|_{\Phi=\pi} &=& - 4\,\pi \,.
\label{appendixF_30}
\end{eqnarray}

\noindent
From Eqs.~(\ref{appendixF_25}) and (\ref{appendixF_30}) we conclude the validity of inequality (\ref{appendixF_5}).

\section{Proof of inequality (\ref{f10_45})}\label{appendixG}

Eq.~(\ref{f10_45}) is given by 
\begin{eqnarray}
\left|\,\sin \Phi\,\frac{z\,\left( z - 1 \right)}{1+z^2-2 z \cos \Phi} - \sin \Phi\,\right| 
\le \frac{2}{1 + z }\,,
\label{appendixG_5}
\end{eqnarray}

\noindent
which is equivalent to the inequality 
\begin{eqnarray}
h_{11} &=& 
\frac{\left(1 - w^2 \right)\;\left(2\,w\,z - z - 1 \right)^2\;\left(1+z\right)^2}{\left(1 + z^2 - 2\,w\,z\right)^2}
- 4 \le 0\,.
\label{appendixG_15}
\end{eqnarray}

\noindent
The extremal conditions $h_{11,z}=0$ and $h_{11,w}=0$ lead to the both relations
\begin{eqnarray}
0 &=& \left(1+z\right)\left(1 - w^2 \right) \left(2\,w\,z - z - 1\right)\left(2\,z^2\,w^2 - 2\,w\,z - z^2 + 1\right),
\label{appendixG_20}
\\
0 &=& \left(1 + z\right)^2\;\left(2\,w\,z - z - 1 \right) 
\nonumber\\
&& \times \left(4\,w^3\,z^2 - 4\,w^2\,z^3 -4\,w^2\,z + w\,z^3 + w\,z^2 + w\,z + w + 2\,z^3 - 2\,z^2 \right).
\label{appendixG_25}
\end{eqnarray}

\noindent
The solutions of (\ref{appendixG_20}) are given by 
\begin{eqnarray}
w_1 &=& - 1\,,
\label{appendixG_30}
\\
w_2 &=& 1\,,
\label{appendixG_35}
\\
w_3 &=& \frac{1+z}{2\,z}\,,
\label{appendixG_40}
\\
w_{4,5} &=& \frac{1 \pm \sqrt{2\,z^2 - 1}}{2\,z}\,,
\label{appendixG_45}
\\
z_1 &=& \frac{1}{2\,w-1}\,,
\label{appendixG_50}
\\
z_{2,3} &=& \frac{w \pm \sqrt{1 - w^2}}{2\,w^2 - 1}\,.
\label{appendixG_55}
\end{eqnarray}

\noindent
Inserting (\ref{appendixG_30}), (\ref{appendixG_35}), (\ref{appendixG_45}) and (\ref{appendixG_55}) into 
(\ref{appendixG_25}) yields 
\begin{eqnarray}
0 &=& \left(7\,z^2 + 3\,z^3 + 5\,z + 1 \right),
\label{appendixG_60}
\\
0 &=& \left( z - 1\right)^4,
\label{appendixG_65}
\\
0 &=& 2\,z^5 + z^4 - 2\,z^3 - 2\,z^2 + 1 \pm \sqrt{2\,z^2 - 1} \left( - z^4 - 2\,z^3 + 2\,z^2 + 2\,z - 1 \right),
\label{appendixG_70}
\\
0 &=& w\,\left( 1 - w^2 \right)\,\left( 1 + w^2 \right) \Bigg[
\left( 8\,w^7 - 16\,w^5 + 12\,w^4 + 2\,w^3 -12\,w^2 + 6\,w - 1 \right) \sqrt{1 - w^2}
\nonumber\\
&& \pm \left( 8\,w^7 - 12\,w^6 - 16\,w^5 +24\,w^4 + 2\,w^3 - 11\,w^2 + 6\,w - 1 \right)\Bigg] \,,
\label{appendixG_75}
\end{eqnarray}

\noindent
while inserting (\ref{appendixG_40}) or (\ref{appendixG_50}) into (\ref{appendixG_25}) yields an identity $0 = 0$.
Obviously, Eq.~(\ref{appendixG_60}) has no real solution for variable $z \ge 0$.
The only solution of Eq.~(\ref{appendixG_65}) is given by 
\begin{eqnarray}
z_1 &=& 1\,.
\label{appendixG_80}
\end{eqnarray}

\noindent
In order to find the solutions of (\ref{appendixG_70}), we first have to bring all those 
terms proportional to $\sqrt{2\,z^2 - 1}$ on the left side, then squaring both sides and obtain 
\begin{eqnarray}
0 &=& \left(1 + z \right)^4\;\left(z - 1 \right)^6\,,
\label{appendixG_85}
\end{eqnarray}

\noindent
according to which it follows that the only solution of (\ref{appendixG_70}) 
is already given by (\ref{appendixG_80}). In order to find the solutions of (\ref{appendixG_75}), 
we first have to bring all those terms proportional to $\sqrt{1 - w^2}$ on the left side, then squaring both sides 
and obtain 
\begin{eqnarray}
0 &=& w^2 \, \left(1 - w^2 \right)\,\left(2 \, w^2 - 1 \right)^6\,.
\label{appendixG_90}
\end{eqnarray}

\noindent
The solutions of Eq.~(\ref{appendixG_90}) are given by (\ref{appendixG_30}), (\ref{appendixG_35}) and 
\begin{eqnarray}
w_{6,7} &=& \pm \frac{1}{\sqrt{2}}\,.
\label{appendixG_100}
\end{eqnarray}

\noindent
Collecting the results (\ref{appendixG_30}) - (\ref{appendixG_45}), (\ref{appendixG_80}) 
and (\ref{appendixG_100}) together, we obtain the extremal points 
\begin{eqnarray}
P_{\rm e1}: \left(z=1 , w = 0\right)\,,
\label{appendixG_105}
\\
P_{\rm e2}: \left(z=1 , w = 1\right)\,.
\label{appendixG_110}
\end{eqnarray}

\noindent
While (\ref{appendixG_110}) is just one point on one of the boundaries, the numerical values of function 
$h_{11}$ (\ref{appendixG_105}) is given by 
\begin{eqnarray}
h_{11}\,\bigg|_{P_{\rm e1}} &=& 0\,.
\label{appendixG_120}
\end{eqnarray}

\noindent
The boundaries are given by 
\begin{eqnarray}
h_{11}\,\bigg|_{z=0} &=& - 3 - w^2 \le 0\,,
\label{appendixG_125}
\\
h_{11}\,\bigg|_{z=\infty} &=& \left( 1 - w^2 \right)\;\left( 2\,w - 1 \right)^2 - 4 \le 0\,,
\label{appendixG_130}
\\
h_{11}\,\bigg|_{w=-1} &=& - 4\,,
\label{appendixG_135}
\\
h_{11}\,\bigg|_{w=1} &=& -4\,.
\label{appendixG_140}
\end{eqnarray}

\noindent
From Eqs.~(\ref{appendixG_120}) - (\ref{appendixG_140}) we conclude the validity 
of inequality (\ref{appendixG_15}) and (\ref{appendixG_5}), respectively. 

\end{document}